\documentclass[conference]{IEEEtran}

\usepackage[space,compress,sort]{cite}
\usepackage{amsmath,amssymb,amsfonts}
\usepackage{color}
\usepackage{algorithm}
\usepackage{algorithmic}
\usepackage[final]{graphicx}

\title{Throughput Maximization for Laser-Powered UAV Wireless Communication Systems}
\author{\IEEEauthorblockN{Jie Ouyang$^*$, Yueling Che$^*$, Jie Xu$^\dagger$, and Kaishun Wu$^*$}
 \IEEEauthorblockA{$^*$School of Computer Science and Software Engineering, Shenzhen University \\
  $^\dagger$School of Information Engineering, Guangdong University of Technology\\
  E-mail: ouyangjie2016@email.szu.edu.cn, yuelingche@szu.edu.cn, jiexu@gdut.edu.cn, wu@szu.edu.cn}}

\begin{document}

\maketitle
\newcommand{\mv}[1]{\mbox{\boldmath{$ #1 $}}}

\begin{abstract}
Laser power has become a viable solution to provide convenient and sustainable energy supply to unmanned aerial vehicles (UAVs). In this paper, we study a laser-powered UAV wireless communication system, where a laser transmitter sends laser beams to charge a fixed-wing UAV in flight, and the UAV uses the harvested laser energy to communicate with a ground station. To maintain the UAV's sustainable operation, its total energy consumption cannot exceed that harvested from the laser transmitter. Under such a laser energy harvesting constraint, we maximize the downlink communication throughput from the UAV to the ground station over a finite time duration, by jointly optimizing the UAV's trajectory and its transmit power allocation. However, due to the complicated UAV energy consumption model, this problem is non-convex and difficult to be solved. To tackle the problem, we first consider a special case with a double-circular UAV trajectory which balances the tradeoff between maximizing the performance of laser energy harvesting versus wireless communication at the UAV. Next, based on the obtained double-circular trajectory, we propose an efficient solution to the general problem, by applying the techniques of alternating optimization and sequential convex programming (SCP). Finally, numerical results are provided to validate the communication throughput performance of the proposed design.
\end{abstract}

  \IEEEpeerreviewmaketitle
\section{Introduction}
The UAV-assisted wireless communication has recently become a promising solution to improve the coverage and network capacity of the terrestrial wireless communication systems, by exploiting UAVs as mobile relays or mobile base stations (see, e.g., \cite{Zeng.UAV.16,BorEl2016,QJR}). However, the UAVs' operation is quite energy-consuming in order to support their propulsion in the air, communication with the ground devices, as well as various application-specific purposes; while conventional battery-powered UAVs only have very limited battery capacity. Therefore, it is  challenging to practically implement the UAV-assisted wireless communication systems in a large scale. In order to alleviate this problem, various approaches have been proposed to reduce the UAV's energy consumption, by, e.g., reducing the UAV's weight \cite{Flight.05} and designing energy efficient UAV traveling path \cite{C2.15}\cite{Zeng.arXivAug.16}. Despite these research efforts, the energy supply for the battery-powered UAVs is still fundamentally unsustainable due to the finite battery capacity.

Recently,  laser power is becoming a viable solution to provide unlimited endurance aloft for UAVs in flight. A laser-powered UAV is installed with photovoltaic receivers to harvest laser power from a laser transmitter. As compared to other wireless power transfer (WPT) techniques enabled by radio frequency (RF) signals (see, e.g.,\cite{Derrick.17,Li.Globe.14,Y.L.15}), the laser-beamed power transfer is able to deliver much larger energy amounts to the receivers with narrower energy beam divergence. For instance, it is demonstrated by the LaserMotive company that several hundred watts can be harvested at the laser power harvester \cite{Killinger.FSO.02}. Therefore,  the laser WPT is expected to  efficiently support various energy-hungry operations of the UAVs over a long distance. The filed tests in \cite{LaserUAV.13} and \cite{Laservehicle.11} have successfully shown the feasibility of laser-powered UAVs.

\begin{figure}
\centering
\DeclareGraphicsExtensions{.eps,.mps,.pdf,.jpg,.png}
\DeclareGraphicsRule{*}{eps}{*}{}
\includegraphics[angle=0,width=0.45\textwidth]{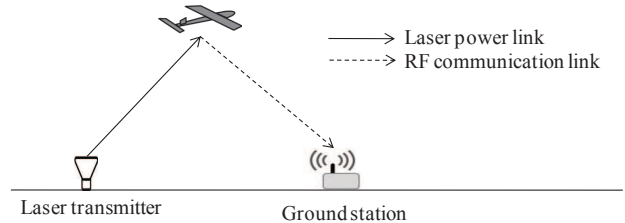}
\caption{Illustration of a laser-powered UAV wireless communication system.}
\label{fig: sys}
\vspace{-2mm}
\end{figure}

In this paper, we study a laser-powered UAV wireless communication system as shown in Fig.~\ref{fig: sys}, where a laser transmitter delivers laser energy to charge a UAV in flight, and the UAV uses the harvested energy from the laser power link to support its flight and downlink RF information transmission to a ground station. This system is a promising solution to provide sustainable UAV wireless communications and is expected to find abundant applications in the era of fifth generation (5G) cellular networks. However, the practical operation of this system faces various design challenges. First, the harvested laser power and the communication rate critically depend on the UAV locations.
When the laser transmitter and the ground station are distributed at different locations, the UAV should carefully design its locations over time (a.k.a. trajectory) to balance the tradeoff between maximizing the performance of laser energy harvesting versus RF wireless communication. Next, in order to maintain the self-sustainable operation, the UAV is subject to a so-called laser energy harvesting constraint, i.e., the UAV's total energy consumption cannot exceed the amount of its harvested energy from the laser transmitter. In addition, the UAV's energy consumption (especially that for propulsion) model is quite complicated, and critically depends on various issues such as the UAV's trajectory, velocity, and acceleration \cite{Zeng.arXivAug.16}. By combining all these issues,  how to optimize the performance of the laser-powered UAV wireless communication system is a very difficult task. This thus motivates our investigation in this work.


In this paper, we particularly focus on maximizing the downlink throughput from the UAV to the ground station over a finite time duration, by jointly optimizing the UAV's trajectory and its transmit power allocation over time, subject to the UAV's laser energy harvesting constraint, and its maximum velocity and acceleration constraints. This problem is shown to be non-convex and thus difficult to be solved. To tackle the problem, we first consider a special case with a double-circular UAV trajectory, where the UAV flies over two circles centered above the laser transmitter and the ground station, respectively, to balance the tradeoff between maximizing the performance of laser energy harvesting versus RF wireless communication at the UAV. In this case, the UAV trajectory design is simplified as the optimization of the UAV's flying velocities and radius over the two circles, which can be efficiently solved. Next, based on the obtained double-circular trajectory, we propose an efficient solution to the general problem, by applying the techniques of alternating optimization and sequential convex programming (SCP). Finally, numerical results show that the proposed design significantly improves the performance of the laser-powered UAV communication system as compared to alternative benchmarks.

In the literature, there have been various prior studies investigating UAV-related wireless communications (see e.g., \cite{Zeng.UAV.16,BorEl2016,QJR} and the references therein) and WPT \cite{xu.17}\cite{xu1.17}.
   However, to our best knowledge, the design of laser-powered UAV wireless communication with laser energy harvesting constraints has not been investigated yet.

\begin{figure*}[ht]
   \begin{small}
   \begin{align*}
   \frac{1}{\eta} \sum_{n=1}^{N} p[n]\delta_t + \sum_{n=1}^{N}\delta_t \left(c_1\|\mv v[n]\|^3 + \frac{c_2}{\|\mv v[n]\|}\left[1+ \frac{\|\mv a[n]\|^2-\frac{(\mv a^T(n)\mv v(n))^2}{\|\mv v[n]\|^2}} {g^2}\right] \right) + \Delta \kappa \leq \sum_{n=1}^{N} \left( \frac{ \delta_t C \varphi e^{-\alpha \sqrt{||\mv q\left[n\right]||^2 + H^2}}}{(D + \sqrt{||\mv q[n]||^2 + H^2} \Delta \theta)^2 } \right) \label{eq: e}
   \tag{11}  
   \end{align*}
   \end{small}
   \hrulefill
   \end{figure*}

  \section{System Model and Problem Formulation}
  We consider a laser-powered fixed-wing UAV wireless communication system as shown in Fig. 1, in which the UAV can collect the laser energy from a laser transmitter, and communicate with the ground station in the RF band.
  We focus on a particular time period with finite duration $T>0$, which is discretized into $N$ time slots each with equal duration $\delta_t$.
  We assume that the laser transmitter is located at the origin $(0,0,0)$ in a three-dimensional (3D) Cartesian coordinate system, and the ground station is located at $(L,0,0)$. 
  Suppose that the UAV flies at a constant altitude $H\!>\!0$ with a time-varying location $(x[n],y[n],H)$ at slot $n \in \{1,\ldots,N\}$. For notational convenience, we denote $\mv q[n]\!=\!(x[n],y[n])$ as the UAV's location projected on the horizontal plane at slot $n \in\{1,\ldots,N\}$, and $\mv \mu=(L,0)$ as that of the ground station, respectively.
  At slot $n$, the distance between the UAV and the laser transmitter is
  $d_b[n] = \sqrt{\|\mv q [n]\|^2 + H^2}$, and that between the UAV and the ground station is $d_s[n] = \sqrt{ \|\mv q[n]-\mv \mu\|^2 + H^2}$.

  First, consider the downlink information transmission over the RF communication link from the UAV to the ground station. By considering the free space path loss model, the channel power gain at slot $n$ is
  \begin{equation}
  h_s[n] = \beta_0 d_s^{-2}[n] = \frac{\beta_0}{||\mv q[n]-\mv \mu||^2 + H^2}, n \in \{1,...,N\}, \nonumber
  \end{equation}
  where $\beta_0$ is the channel power gain at a reference distance $d_0=1$ meter (m). The instantaneous downlink throughput at slot $n$ (in bps/Hz) is
  \begin{align}
  R[n] &= \log_{2}\left(1 + \frac{p[n] \gamma}{||\mv q[n]\!-\!\mv \mu||^2 \!+\! H^2}\right), n \in \{1,...,N\}, \label{eq: Rate}
  \end{align}
  where $p[n]\ge 0$ is the UAV's dowlink transmit power to the ground station at slot $n$, $\sigma^2$ denotes the noise power, and $\gamma= \beta_0/ \sigma^2$ represents the reference signal-to-noise ratio (SNR). As a result, the cumulative downlink throughput achieved over all $N$ time slots is expressed as
  \begin{equation}
  R_\text{sum} = \sum_{n=1}^{N}\delta_t R[n].\label{eq: Rsum}
  \end{equation}

  We consider the transmission related energy consumption as the dominant energy consumption at the UAV for its downlink communication, and ignore other terms caused by, e.g., circuits in the RF chain and baseband signal processing. Hence, the total energy consumption at the UAV for its downlink communication over all $N$ slots can be expressed as
  \begin{equation}
  P_m = \frac{1}{\eta} \sum_{n=1}^{N} p[n] \delta_t, \label{eq: P_m}
  \end{equation}
  where $0 < \eta\le 1$ denotes the RF chain efficiency.

  In addition to the communication related energy consumption, the UAV also needs to consume energy for propulsion.
  The total amount of propulsion energy over all $N$ slots, denoted by $P_f$, can be expressed as follows based on the analytic energy consumption model in \cite{Zeng.arXivAug.16}.
  \begin{align}
  P_f \!=\! \sum_{n=1}^{N} \delta_t &\!\left[\!c_1\|\mv v[n]\|^3 \!+\! \frac{c_2}{\|\mv v[n]\|}\!\left(\!1 \!+\! \frac{\|\mv a[n]\|^2 \!\!-\!\! \frac{(\mv a^T[n]\mv v[n])^2}{\|\mv v[n]\|^2}} {g^2}\!\right)\!\!\right]\! \nonumber \\
  &+  \frac{1}{2}m(\|\mv v[N]\|^2 - \|\mv v[1]\|^2), \label{eq: P_f}
  \end{align}
  where $c_1$ and $c_2$ are two parameters related to the UAV's weight, wing area, air density, etc., $g$ is the gravitational acceleration with nominal value (9.8 m/s$^2$), $m$ is the mass of the UAV including all its payload, and $\mv v[n]$ and $\mv a[n]$ are the UAV's velocity and its acceleration at slot $n$, respectively. It should be noted that the UAV trajectory and velocity over slots can be updated according to the following equations:
  \begin{align}
  \mv v[n+1] &= \mv v[n] \!+\! \mv a[n]\delta_t,  \ \ \forall  n \in \{1,...,N\!-\!1\}, \label{eq: vlink}  \\
  \mv q[n+1] &= \mv q[n] \!+\! \mv v[n]\delta_t \!+\! \frac{1}{2} \mv a[n] \delta_t^2, \forall n \in \{1,...,N\!-\!1\}.  \label{eq: qlink}
  \end{align}

  By combining the energy consumption $P_m$ for communication and $P_f$ for propulsion, the total amount of energy consumption at the UAV over all $N$ slots is expressed as
  \begin{equation}
  P_c = P_m + P_f.  \label{eq:P_mf}
  \end{equation}

  Next, consider the UAV's energy harvesting over the laser power link. We assume that the laser transmitter adopts a fixed transmit power $\varphi>0$.
  Accordingly, the received signal strength at the UAV at slot $n$ is expressed as \cite{Killinger.FSO.02}
  \begin{equation}
  P_s[n] = \delta_t \varphi \frac{A}{(D \!+\! d_b[n] \Delta \theta)^2}  \chi e^{-\alpha d_b[n]}, n \in \{1,...,N\}, \label{eq:P_h1}
  \end{equation}
  where $A$ is the area of the receiver telescope or collection lens, $D$ is the size of the initial laser beam, $\Delta \theta$ is the angular spread, $\chi$ is the combined transmission receiver optical efficiency, and $\alpha$ is the attenuation coefficient of the medium in m$^{-1}$. By considering a linear energy harvesting model with a constant laser energy harvesting efficiency $\omega\in(0,1)$, the amount of harvested laser energy at the UAV at slot $n$ is given by
  \begin{equation}
  P_h[n]=\omega P_s[n].  \label{eq:P_h11}
  \end{equation}
  It is noted that in (\ref{eq:P_h1}) and (\ref{eq:P_h11}) under clear weather conditions, $\alpha$ is of a very small value with $10^{-6}$m. Hence, the variations of $P_s[n]$ and thus $P_h[n]$ over the distance $d_b[n]$ are dominated by $(D + d_b[n] \Delta \theta)^{-2}$ in this case.  Moreover, notice that  $\Delta \theta$ is normally very small and the laser transmit power $\varphi$ is large (e.g., $\Delta \theta = 3.4 \times 10^{-5}$ and $\varphi = 1$~kw \cite{Hemani.FSO} ); therefore, $P_h[n]$ generally decreases much slower over the distance $d_b[n]$ as compared to the case of RF energy harvesting \cite{Derrick.17}. This also explains that the laser power can have a much longer charging distance to support energy-demanding applications such as UAVs.
  By letting $C = \omega A \chi$, the total harvested laser energy at the UAV over all $N$ slots is obtained as
  \begin{equation}
  \tilde P_h = \sum_{n=1}^{N} P_h[n] =\sum_{n=1}^{N} \frac{ \delta_t C \varphi e^{-\alpha d_b[n]}}{(D + d_b[n] \Delta \theta)^2 }.  \label{eq:P_h}
  \end{equation}

  To provide unlimited endurance aloft, the energy consumption at the UAV in (\ref{eq:P_mf}) cannot exceed the amount of its harvested laser energy in (\ref{eq:P_h}). Accordingly, we have the laser energy harvesting constraint as $P_c \leq \tilde P_h$, which is further explicitly expressed in (\ref{eq: e}), with $\Delta \kappa = \frac{1}{2}m(\|\mv v[N]\|^2 - \|\mv v[1]\|^2)$ denoting the UAV's kinetic energy.

\begin{tabular}{rl}
\end{tabular}
   Our objective is to maximize the cumulative downlink throughput $R_\text{sum}$ in (\ref{eq: Rsum}) by jointly optimizing the UAV's trajectory $\{ \mv q[n]\}$, the associated velocity $\{\mv v[n]\}$, the acceleration $\{ \mv a[n]\}$, and its transmit power $\{ p[n]\}$ over the $N$ time slots. Therefore, the problem is mathematically formulated as:
   \begin{align}
  \text{(P1):}\max_{\substack{\{ \mv q[n],p[n],\\\mv v[n],\mv a[n]\}}} ~&  \sum_{n=1}^{N} \delta_t \log_{2}\left(1 + \frac{p\left[n\right] \gamma}{\left[||\mv q[n]-\mv \mu||^2 + H^2\right]}\right) \nonumber \\
   \mathrm{s.t.}~& \setcounter{equation}{11} \| \mv a[n]\| \leq a_{\max}  , \forall n \in \{1,...,N\},\label{eq: alimit} \\
   & \| \mv v[n]\| \leq V_{\max} , \forall n \in \{1,...,N\},\label{eq: vlimit} \\
   & p[n] \geq 0   , \forall n \in \{1,...,N\}, \label{eq: P0}  \\
   & (\ref{eq: vlink}),~(\ref{eq: qlink}),~\text{and}~(\ref{eq: e}),  \nonumber
  \end{align}
   where $V_{\max}$ and $a_{\max}$ represent the maximum  allowable velocity and acceleration for the UAV, respectively. Note that (P1) is a non-convex optimization problem, as the objective function in (P1) and the laser energy harvesting constraint in (\ref{eq: e}) are both non-convex. Therefore, problem (P1) is generally a very difficult problem to be solved optimally. To tackle this problem, in Section III we first study a special case with a double-circular UAV trajectory. Based on the solution obtained from the special case, we then propose an efficient algorithm to solve problem (P1) in Section IV.

  \section{Solutions to (P1) under a Double-Circular Trajectory}
\begin{figure}
\centering
\DeclareGraphicsExtensions{.eps,.mps,.pdf,.jpg,.png}
\DeclareGraphicsRule{*}{eps}{*}{}
\includegraphics[angle=0,width=0.38\textwidth]{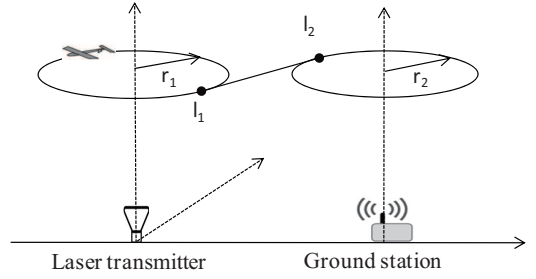}
\caption{Illustration of the double-circular UAV trajectory.}
\label{fig: circle}
\vspace{-2mm}
\end{figure}

   In order to properly balance the tradeoff between maximizing the performance of laser energy harvesting and wireless communication, this section considers a special double-circular UAV trajectory as shown in Fig.~\ref{fig: circle}, where the UAV flies over two circles centered above the laser transmitter and the ground station for efficient laser energy harvesting and RF wireless communication, respectively. In this design, we consider fixed downlink power allocation $p$ at the UAV for wireless communication.
   Specifically, in the first circle centered at the laser transmitter of radius $r_1$, the UAV flies with a constant velocity $V_1$ for a total of $n_1\geq0$ laps, and aims to maximize the amount of the net harvested energy, which is defined as the amount of the UAV's harvested laser energy offset by its energy consumption for propulsion. In the second circle centered at the ground station of radius $r_2$, the UAV flies with a constant velocity $V_2$ for a total of $n_2\geq0$ laps, and aims to maximize the energy efficiency for the UAV's downlink RF wireless communication, which is defined as the number of delivered bits per unit energy. The two circles are connected by a cotangent line $l_1l_2$, over which the UAV flies with a constant acceleration $\mv a_{12}$. In the following, we first design the optimal UAV velocity and flight radius over each circle, and then derive the optimal number of flight laps for both circles.

  First, we consider the circle for UAV laser energy harvesting. Since the UAV adopts the same constant velocity in each slot with $\|\mv v[n]\|=V_1$, the acceleration $\mv a[n]$ is perpendicular to the velocity with $\mv a[n]^T \mv v[n] = 0$, and thus $\|\mv a[n]\| = V_1^2/r_1$. Thus, the UAV's energy consumption for propulsion in (\ref{eq: P_f}) in each slot is obtained as
  \vspace{-1mm}
  \begin{equation}
  P_f^\prime = \delta_t\left(c_1 + \frac{c_2}{g^2r_1^2}\right)V_1^3 + \delta_t\frac{c_2}{V_1}. \label{eq: pfp}
  \vspace{-1mm}
  \end{equation}
  The harvested laser energy at the UAV from the laser transmitter in (\ref{eq:P_h}) in each slot is obtained as
  \vspace{-1mm}
  \begin{equation}
  P_h^\prime = \frac{\delta_t C \varphi e^{-\alpha \sqrt{H^2+r_1^2}}}{(D + \sqrt{H^2+r_1^2} \Delta \theta)^2}. \label{eq: php}
  \vspace{-1mm}
  \end{equation}
  Since the UAV's harvested laser energy in (\ref{eq: php}) and consumed energy for propulsion in (\ref{eq: pfp}) for each slot are both constants, for a given number of laps $n_1$ over the circle above the laser transmitter, the UAV's net harvested energy over this circle can be maximized by maximizing $(P_h' - P_f')/\delta_t$, as formulated in the following problem.
  \vspace{-1mm}
  \begin{equation}
  \text{(P2)}:\max_{\substack{0\leq V_1 \leq V_{\max},\\r_1 \geq 0 }} \frac{C \varphi e^{-\alpha\sqrt{H^2+r_1^2}}}{(D \!+\!\sqrt{H^2\!+\!r_1^2} \Delta \theta)^2} \!-   \!\left(\!c_1 \!\!+\!\! \frac{c_2}{g^2r_1^2}\!\right)\!V_1^3 \!-\! \frac{c_2}{V_1}. \label{eq: net energy} \nonumber
  \vspace{-1mm}
  \end{equation}

  Note that under a given $r_1$, the objective function of problem (P2) is convex with respect to the velocity $V_1$. By checking its first-order derivative, the optimal UAV velocity $V_1^*$ can be obtained as
  \begin{equation}
  V_1^*(r_1)= \min\left( \left(\frac{c_2}{3(c_1+c_2/(g^2r_1^2))}\right)^\frac{1}{4} ,V_{\max}\right). \label{eq: V^*}
  \end{equation}
  By substituting $V_1^*(r_1)$ in (\ref{eq: V^*}) into (P2), (P2) is transformed into an univariate optimization problem of radius $r_1$.
   Then we adopt a one-dimensional exhaustive search to find the optimal solution $r_1^*$. As a result, the optimal solution to (P2) is obtained as $r_1^*$ and we have $V_1^*=V_1^*(r_1^*)$.

  Next, we consider the circle centered above the ground station, where the UAV aims to maximize the energy efficiency for its RF wireless communication. Similarly as for solving (P2), it can be shown that under a given radius $r_2$, the optimal velocity $V_2^*$ in each slot can be obtained as $V_2^*=V_1^*(r_2)$ in (\ref{eq: V^*}). According to \cite{Zeng.arXivAug.16}, by using the optimal $V_2^*$, the energy efficiency of the UAV can be expressed as
 \begin{equation}
 \vartheta(r_2) = \frac{\log_2(1+\frac{p\gamma}{r_2^2 + H^2})}{( c_1 + c_2/(g^2 r_2^2)) {V_2^*}^3 - c_2/V_2^*}. \label{eq: Efmax}
 \end{equation}
 By using the approximation $\ln(1+x) \approx x$ if $x \ll 1$, the optimal radius $r_2^*$ that maximizes the energy efficiency in (\ref{eq: Efmax}) is obtained as
 \begin{equation}
 r_2^* = V_2^*\sqrt{\frac{Hc_2^{1/2}}{g(c_1{V_2^*}^4-c_2)^{1/2}}},
 \end{equation}
  Therefore, the UAV's flying radius and velocity over this circle are obtained as $r_2^*$ and $V_2^* = V_1^*(r_2^*)$.

  Furthermore, for the straight-and-level trajectory from $l_1$ to $l_2$, the adopted constant acceleration at the UAV is $ \|\mv a_{12}\|=\frac{V_2^{*2}-V_1^{*2}}{2l_{12}}$, where $l_{12}$ denotes the length of the cotangent line $l_1l_2$ and is given as $l_{12}=\sqrt{L^2-(r_1+r_2)^2}$ based on the Pythagorean Theorem.

 Finally, due to the finite time duration under consideration, the UAV's flight laps over the two circles are found to satisfy the following equation:
\begin{equation}
\frac{2 \pi r^*_1 n_1}{V^*_1} + \frac{|V^*_2 - V^*_1|}{\|\mv a_{12}\|} + \frac{2 \pi r^*_2 n_2}{V^*_2}=T.   \label{eq: time}
\end{equation}
Hence, for each given $n_1$ and $n_2$, we can obtain a double-circular trajectory with $\{\mv q[n]\}_{n=1}^N$, velocities $\{\mv v[n]\}_{n=1}^N$ and accelerations $\{\mv a[n]\}_{n=1}^N$ by using the optimal $r_1^*$, $V_1^*$, $r_2^*$, and $V_2^*$, base on which the UAV's total amount of harvested energy $\tilde P_h$ in (\ref{eq:P_h}) and the total amount of consumed energy $P_f$ in (\ref{eq: P_f}) over the designed trajectory $\{\mv q[n]\}_{n=1}^N$ can be easily obtained.
Therefore, the UAV's transmit power is obtained as $p = \eta(\tilde P_h - P_f)/T$. The cumulative downlink rate $R_\text{sum}$ in (\ref{eq: Rsum}) can thus be obtained by substituting $p$ into (\ref{eq: Rate}) and (\ref{eq: Rsum}). By exhaustively searching over all $n_1$ and $n_2$ satisfying (\ref{eq: time}), the optimal $n_1^*$ and $n_2^*$ are obtained to maximize $R_\text{sum}$. Therefore, with the optimal $n_1^*$, $n_2^*$, $r_1^*$, $V_1^*$, $r_2^*$, and $V_2^*$, the optimal double-circular trajectory is obtained.

  \section{Alternative Trajectory and Power Allocation Optimization for (P1)}
   In this section, based on the double-circular trajectory obtained in Section III, we address problem (P1) by alternatively optimizing the UAV's transmit power and trajectory. In the following, we first optimize the UAV's transmit power under a given trajectory, and then optimize the UAV's trajectory under a given power allocation. At last, we propose an efficient iterative algorithm to solve problem (P1).

 \subsection{Optimal Power Allocation Under Given Trajectory}
 This subsection maximizes the UAV's downlink throughput by optimizing $\{p[n]\}$ under given trajectory $\{\mv q[n]\}$, $\{\mv v[n]\}$, and $\{\mv a[n]\}$. In this case, problem (P1) is reduced to the following convex optimization problem.
 \begin{align}
  \text{(P1.1):}\max_{\substack{p[n]}} ~& \sum_{n=1}^{N}\delta_t \log_{2}\left(1 + \frac{p\left[n\right] \gamma}{\left[||\mv q\left[n\right]-\mv \mu||^2 + H^2\right]}\right) \nonumber \\
  \mathrm{s.t.}~& (\ref{eq: e})~\text{and}~(\ref{eq: P0}). \nonumber
  \end{align}
By using the Karush-Kuhn-Tucker (KKT) condition, the optimal power allocation $p^*[n]$ follows the water-filling structure as
\begin{equation}
p^*[n]=\left[\lambda-\frac{||\mv q[n]-\mv \mu||^2 + H^2}{\gamma} \right]^+,	   \label{eq: p^*}
\end{equation}
where $[a]^+ \triangleq \max \left\{a,0\right\}$, and the water level $\lambda$ is chosen such that the constraint in (\ref{eq: e}) is met with equality. It is observed from (\ref{eq: p^*}) that the optimal power allocation for problem (P1) is trajectory-aware, and its value generally increases as the UAV flies towards the  ground station to achieve high downlink throughput.

\subsection{Trajectory Optimization Under Given Power Allocation}
This subsection optimizes the UAV's trajectory $\{\mv q[n]\}_{n=1}^N$ under any given power allocation $\{p[n]\}_{n=1}^N$. In this case, problem (P1) is reduced to
\begin{align}
 \text{(P1.2):} \max_{\substack{\{\mv q[n]\}\\\{\mv v[n]\},\{\mv a[n]\}}} ~& \sum_{n=1}^{N} \delta_t \log_{2}\left(1 + \frac{p\left[n\right] \gamma}{\left[||\mv q[n]-\mv \mu||^2 + H^2\right]}\right) \nonumber \\
 \mathrm{s.t.}~& (\ref{eq: vlink}),~(\ref{eq: qlink}),~(\ref{eq: e}),~(\ref{eq: alimit})~\text{and}~(\ref{eq: vlimit}).  \nonumber
 \end{align}

Note that the objective function of (P1.2) and the constraint in (\ref{eq: e}) are both non-convex. Therefore, problem (P1.2) cannot be solved by standard convex optimization techniques. To facilitate the derivation, we use an upper bound $P_{ub}$ to replace $P_f$ in \eqref{eq: P_f} for calculating the UAV's propulsion energy, which is given as
  \begin{equation}
   P_f \leq \sum_{n=1}^{N}\delta_t \!\left[\!c_1\|\mv v[n]\|^3 \!+\! \frac{c_2}{\|\mv v[n]\|}\left(1\!+\! \frac{\|\mv a[n]\|^2} {g^2}\right)\!\right]\! \!+\! \Delta \kappa \triangleq P_{ub}. \nonumber
   \end{equation}
The upper bound is tight for the constant-speed flight, in which case we have $\mv a[n]^T\mv v[n]=0$ at any slot $n$.
After using $P_{ub}$ to replace $P_{f}$, we also introduce slack variables $\zeta_n$ and ${\tau_n}$ and reformulate (P1.2) as follows:
 \begin{align}
 \text{(P1.2-1):}& \max_{\substack{\{\mv q[n],\zeta_n\\\mv v[n],\mv a[n],\tau_n\}}} \sum_{n=1}^{N} \delta_t \log_{2}\left(1 + \frac{p[n] \gamma}{\left[\|\mv q[n]-\mv \mu\|^2 + H^2\right]}\right) \nonumber \\
 \mathrm{s.t.}~& \sum_{n=1}^{N}\delta_t\left(c_1\|\mv v[n]\|^3 + \frac{c_2}{\tau_n} + \frac{c_2 \|\mv a[n]\|^2} {g^2 \tau_n} \right) + \Delta \kappa \nonumber  \\
 & + \frac{1}{\eta} \sum_{n=1}^{N} p[n]\delta_t \leq \sum_{n=1}^{N} \left( \frac{ \delta_t C \varphi e^{-\alpha \sqrt{\zeta_n}}}{(D_1 + \sqrt{\zeta_n} \Delta \theta_1)^2 } \right),   \label{eq: pup1}  \\
 & \zeta_n \geq 0, \forall n\in\{1,\ldots,N\}, \label{eq: zetag0} \\
 & \tau_n \geq 0, \forall n\in\{1,\ldots,N\}, \label{eq: tau1} \\
 & ||\mv q[n]||^2 + H^2 \leq \zeta_n, \forall n\in\{1,\ldots,N\},\label{eq: zetaq}\\
 & \| \mv v[n]\|^2 \geq \tau_n^2, \forall n\in\{1,\ldots,N\}, \label{eq: tau} \\
 & (\ref{eq: vlink}),~(\ref{eq: qlink}),~(\ref{eq: alimit})~\text{and}~(\ref{eq: vlimit}).  \nonumber
 \end{align}
 It can be shown that at the optimal solution to (P1.2-1), we must have $\zeta_n= ||\mv q\left[n\right]||^2 + H^2$ and $\tau_n = \| \mv v[n]\|$, since otherwise one can always decrease $\zeta_n$ or increase $\tau_n$ to increase the objective value without violating the constraints.

 In the following, we use the technique of SCP to approximate the non-convex objective function and constraints in (P1.2-1) into convex terms. First, consider the non-convex constraint (\ref{eq: tau}), in the left-hand-side of which $\|\mv v[n]\|^2$ is convex and differentiable with respect to $\mv v[n]$.
For any given point $\left\{ \mv v_i[n] \right\}$, we have
\begin{align}
 ||\mv v[n]||^2 & \! \geq \! ||\mv v_i[n]||^2\! + \!2 \mv v_i^T[n](\mv v[n]\!-\!\mv v_i[n]) \!\triangleq\! \psi_{lb}(\mv v[n]), \label{eq: vj}
\end{align}
where the equality holds at the point $\mv v[n]= \mv v_i[n]$.
The constraint in (\ref{eq: tau}) can thus  be approximated as
the following convex constraint, in which $\psi_{lb}(\mv v[n])$ is linear with respect to $\mv v[n]$.
\begin{equation}
\psi_{lb}(\mv v[n]) \geq \tau_n^2, \label{eq: vtau}
\end{equation}

Next, consider the constraint in (\ref{eq: pup1}). First, by using (\ref{eq: vj}), the change of the UAV's kinetic energy $\Delta \kappa$ can also be approximated  as a convex term. With such approximation, the harvested energy $P_h(\zeta_{n})$ and the propulsion energy $P_{ub}$ in (P1.2-1) are both jointly convex with respect to $\left\{\zeta_n \right\}$ and $\left\{\mv v[n], \mv a[n], \tau_n\right\}$. respectively. Thus, for any given local point $\left\{\zeta_{in} \right\}$, we replace (\ref{eq: pup1}) with
 \begin{align}
P_{ub} + P_m - P_h(\zeta_{in})- \triangledown P_h(\zeta_{in})^T (\zeta_n-\zeta_{in}) \leq 0, \label{eq: pup2}
\end{align}
where the equality holds at the point $\zeta_n= \zeta_{in}$.

In addition, for the non-concave objective function, we can have its lower bound as follows based on the Taylor expansion
\begin{equation}
R_{lb}(\mv q[n]) \!=\! \sum_{n=1}^{N} \left[\alpha_i[n] \!-\! \beta_i[n](||\mv q[n]\!-\!\mv \mu||^2 \!-\! ||\mv q_i[n] \!-\!\mv \mu||^2)\right], \nonumber
\end{equation}
where
\begin{align}
\alpha_i[n] &= \log_2 \left( 1 + \frac{p[n] \gamma}{\left[||\mv q_i[n] -\mv \mu||^2 + H^2\right]} \right), \nonumber \\
\beta_i[n] &= \frac{(\log_2e) p[n] \gamma}{(p[n] \gamma \!+\! \|\mv q_i[n]\! -\!\mv \mu\|^2 \!+\! H^2)(\|\mv q_i[n] \!-\!\mv \mu\|^2 \!+\! H^2)}. \nonumber
\end{align}
Note that $R_{lb}(\mv q[n])$ is a concave function with respect to $\mv q[n]$. We have
\begin{equation}
 \sum_{n=1}^{N}\log_{2}\left(1 + \frac{p[n] \gamma}{\left[||\mv q[n]-\mv \mu||^2 + H^2\right]}\right) \geq R_{lb}(\mv q[n]),
\end{equation}
where the equality holds at the point $\mv q[n] = \mv q_i[n]$.

By using (\ref{eq: vtau}) and (\ref{eq: pup2}), problem (P1.2-1) is approximated as the following convex problem:
\begin{align}
  \text{(P1.2-2):}\max_{\substack{\{\mv q[n],\zeta_n\\\mv v[n],\mv a[n],\tau_n\}}} ~& \delta_t R_{lb}(\mv q[n]) \nonumber \\
   \mathrm{s.t.}~& (\ref{eq: vlink}),~(\ref{eq: qlink}),~(\ref{eq: alimit}),~(\ref{eq: vlimit}),~(\ref{eq: zetag0}),\nonumber \\
   & (\ref{eq: tau1}),~(\ref{eq: zetaq}),~(\ref{eq: vtau})~\text{and}~(\ref{eq: pup2}). \nonumber
\end{align}
Suppose that the obtained solution as $\{ \mv q_i[n],\mv v_i[n],\mv a_i[n],$ $ \zeta_{in},\tau_{in} \}$ at the $i$-th iteration, we do the approximation as $\{ \mv q_{i+1}[n],\mv v_{i+1}[n],\mv a_{i+1}[n],\zeta_{{(i+1)}n},\tau_{{(i+1)}n} \}$. In the $(i+1)$-th iteration, we solve (P1.2-2). Thus, the original non-convex problem (P1.2) can be approximately solved iteratively.

\subsection{Complete Algorithm for Solving (P1)}

By combining the results in the above two subsections, we use the alternative optimization to efficiently solve the non-convex problem (P1) by solving (P1.1) and (P1.2) iteratively. In particular, we use the double-circular trajectory and the associated equal power allocation as the initial point in the iteration. Therefore, the algorithm for solving (P1) is finally obtained.

\section{Numerical Results}
In this section, we provide numerical results to validate our proposed design. We assume that the laser transmitter and the ground station are separated with distance $L=500$~m, and the altitude of the UAV is fixed to be $H=100$~m. For the laser-powered UAV wireless communication system, the reference SNR is $\gamma=20$~dB, and the maximum UAV velocity and acceleration are assumed to be $V_{\max} = 60$~m/s and $a_{\max} = 6$~m/s$^2$. According to \cite{Zeng.arXivAug.16},\cite{Hemani.FSO}, and \cite{Filippone.flight}, we set $C=0.004$~m$^2$, $\alpha=10^{-6}$~m, $D=0.1$~m, $\Delta \theta=3.4\times10^{-5}$, $\eta=1$, $c_1=9.26\times10^{-4}$~kg/m, and $c_2=2250$~kg$\cdot$m$^3$/s$^4$. We also set $T=100$~s.

\begin{figure}
\centering
\DeclareGraphicsExtensions{.eps,.mps,.pdf,.jpg,.png}
\DeclareGraphicsRule{*}{eps}{*}{}
\includegraphics[angle=0,width=0.30\textwidth]{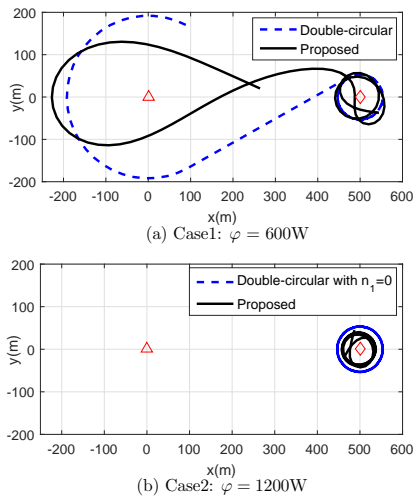}
\caption{The UAV trajectory under different laser transmit power values. The triangle and the diamond denote the laser transmitter and ground station, respectively.}
\label{fig: trajectory}
\vspace{-2mm}
\end{figure}
Fig.~\ref{fig: trajectory}(a) shows the proposed UAV trajectory using the alternating optimization algorithm in Section IV with $\varphi=600$~W. For the initial double-circular trajectory, we obtain a larger velocity $V_1^*=26.43$~m/s with a smaller number of laps  $n_1=0.68$ for the circle centered above the laser transmitter, and with a smaller velocity $V_2^*=17.11$~m/s with a larger number of laps $n_2=2.49$ for the circle centered above the ground station, to balance the tradeoff between laser energy harvesting maximization
versus RF wireless communication quality maximization.
For the proposed trajectory, the UAV flies in a droplet-shaped trajectory closer to the laser transmitter to harvest more energy, and then hovers above the ground station to maintain better wireless communication channels.
Fig.~\ref{fig: trajectory}(b) shows the obtained UAV trajectory with $\varphi=1200$~W. As compared to that in Fig.~\ref{fig: trajectory}(a), the initial double-circular trajectory is reduced to be a single circle above the ground station with $n_1=0$, due to the large amount of harvested laser energy at the UAV from the high-power laser beam, such that the UAV can harvest sufficient amount of laser energy to support its operations by flying on the energy efficiency circle alone. It is also observed that the proposed trajectory is almost of a circular shape with a radius smaller than the initial  circle to improve the downlink communication throughput.

\begin{figure}
\centering
\DeclareGraphicsExtensions{.eps,.mps,.pdf,.jpg,.png}
\DeclareGraphicsRule{*}{eps}{*}{}
\includegraphics[angle=0,width=0.32\textwidth]{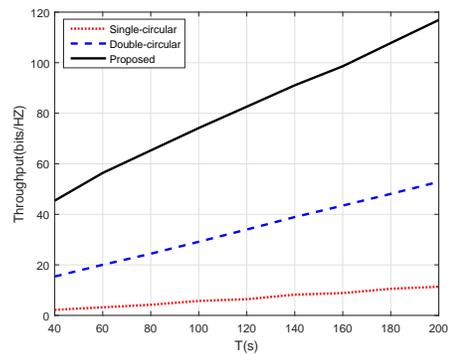}
\caption{Throughput of the proposed system.}
\label{fig: throu}
\vspace{-1mm}
\end{figure}

Fig.~\ref{fig: throu} shows the achievable downlink sum throughput $R_{\text{sum}}$ in bits/Hz versus the UAV's flight time $T$. We compare the proposed joint trajectory and power optimization with two benchmark schemes, i.e., the double-circular trajectory with equal power allocation in Section III, and the single-circular UAV trajectory. In the single-circular UAV trajectory design, the UAV flies over a circle centered above the laser transmitter to maximize the amount of net harvested energy and equally allocates the net harvested energy to each time slot as the UAV's transmit power for its downlink communication. In all three schemes, we set $\varphi=600$~W. The throughputs of all three schemes in Fig.~\ref{fig: throu} increase over time horizon $T$, as expected.
 It is observed that the achievable downlink throughput by the double-circular trajectory outperforms that by the single-circular trajectory. This is because the double-circular trajectory can adaptively fly above both laser transmitter and ground station to efficiently balance the tradeoff between laser energy harvesting versus RF wireless communication.
 It is also observed that the throughput of the proposed design increases significantly over time horizon $T$ as compared to both benchmarks.
 This validates the significance of such joint optimization.

\section{Conclusion}

This paper proposed a new laser-powered UAV wireless communication system. Our objective was to maximize the UAV's cumulative downlink throughput over a finite time duration by jointly optimizing the UAV's trajectory and its transmit power allocation. We proposed an efficient design that alternatively optimizes the UAV trajectory and the power allocation over time. Numerical results under practical system setups validated the performance of the proposed algorithm.


\section*{Acknowledgement}
This work was supported by the National Science Foundation of China (61601308), and the Guangdong Provincial Science and Technology Development Special Fund project (2017A010101033).

  \end{document}